\documentclass[usenatbib]{emulateapj}
\usepackage{amsmath}

\newcommand{\be}{\begin{displaymath}}
\newcommand{\ee}{\end{displaymath}}

\def\lsim{\hbox{\rlap{\raise 0.425ex\hbox{$<$}}\lower 0.65ex\hbox{$\sim$}}}
\def\gsim{\hbox{\rlap{\raise 0.425ex\hbox{$>$}}\lower 0.65ex\hbox{$\sim$}}}

\shorttitle{Interacting hydrogen in Type Ib SNe}
\shortauthors{Mauerhan et al.}

\begin{document}

\title{Stripped-envelope supernova SN\,2004dk is now interacting with hydrogen-rich circumstellar material}

\author{Jon C. Mauerhan\altaffilmark{1,2},        
Alexei V. Filippenko\altaffilmark{2,3}, WeiKang Zheng\altaffilmark{2}, Thomas Brink\altaffilmark{2},
Melissa L. Graham\altaffilmark{4}, Isaac Shivvers\altaffilmark{2},
and Kelsey Clubb\altaffilmark{2} }

\altaffiltext{1}{The Aerospace Corporation, 2310 E. El Segundo Blvd., El Segundo, CA 90245, USA}
\altaffiltext{2}{Department of Astronomy, University of California, Berkeley, CA 94720-3411, USA}
\altaffiltext{3}{Miller Senior Fellow, Miller Institute for Basic Research in Science, University of California, Berkeley, CA 94720, USA}
\altaffiltext{4}{Department of Astronomy, University of Washington, Box 351580, Seattle, WA 98195-1580, USA}

\begin{abstract} The dominant mechanism and time scales over which stripped-envelope supernovae (SNe) progenitor stars shed their hydrogen envelopes are uncertain. Observations of Type Ib and Ic SNe at late phases could reveal the optical signatures of interaction with distant circumstellar material (CSM) providing important clues on the origin of the necessary pre-SN mass loss. We report deep late-time optical spectroscopy of the Type Ib explosion SN\,2004dk 4684 days (13 years) after discovery. Prominent intermediate-width H$\alpha$ emission is detected, signaling that the SN blast wave has caught up with the hydrogen-rich CSM lost by the progenitor system.  The line luminosity is the highest ever reported for a SN at this late stage. Prominent emission features of He\,{\sc i}, Fe, and Ca are also detected. The spectral characteristics are consistent with CSM energized by the forward shock, and resemble the late-time spectra of the persistently interacting Type IIn SNe\,2005ip and 1988Z. We suggest that the onset of interaction with H-rich CSM was associated with a previously reported radio rebrightening at $\sim1700$ days. The data indicate that the mode of pre-SN mass loss was a relatively slow dense wind that persisted millennia before the SN, followed by a short-lived Wolf-Rayet phase that preceded core-collapse and created a cavity within an extended distribution of CSM. We also present new spectra of SNe\,2014C, PTF11iqb, and 2009ip, all of which also exhibit continued interaction with extended CSM distributions.

\end{abstract}


\section{Introduction}

Stripped-envelope supernovae (SNe) are core-collapse explosions that stem from massive stars which have lost most or all of their hydrogen (Types IIb, Ib) and most or all of their helium envelopes (Type Ic). The mechanism by which the progenitors shed this material is uncertain. Possibilities include strong sustained winds from stars evolving into the Wolf-Rayet (WR) phase (Begelman \& Sarazin 1986), discrete super-Eddington eruptions from luminous blue variable (LBV) stars (Smith \& Owocki 2006; Smith 2014), Roche-lobe overflow and accretion onto a binary companion (Wheeler \& Levreault 1985), or common-envelope ejection from a binary system (Podsiadlowski et al. 1992). Since massive stars are typically in close binaries that are likely to interact at least once during their lives (Sana et al. 2012), various combinations of sustained, discrete, and binary-driven mass loss could contribute to evolution of stripped-envelope SN progenitors.

Signatures of interaction with circumstellar material (CSM) have been observed for a subset of stripped-envelope SNe, and over a wide variety of time scales with respect to the explosion time. SNe~Ibn, for example, are events which exhibit namesake relatively narrow emission lines of helium soon after exploding, which implies that the dense CSM is close to the star and was produced by the progenitor just months to several years before core collapse, either from an extreme stellar wind or from a discrete mass ejection (Foley et al. 2007; Shivvers et al. 2017). Other stripped-envelope SNe have exhibited delayed interaction with CSM. For example, SN\,2014C was a seemingly normal hydrogen-poor Type Ib event at first, but subsequently underwent a delayed metamorphosis into a strongly interacting H-rich SN~IIn over the course of a year (Milisavljevic et al. 2015; Margutti et al. 2017). This change suggests that the progenitor either experienced a mass-loss transition, possibly the result of a fast WR wind developing and carving out a cavity within the slower wind from a earlier red-supergiant or LBV phase (Dwarkadas 2005, 2007); a shell ejection from a discrete eruption; or CSM confined by an external radiation field produced by neighboring massive stars (Mackey et al. 2014).

SN\,2004dk was a stripped-envelope SN initially classified as a He-poor Type Ic explosion (Patat et al. 2004), but it eventually exhibited prominent helium lines consistent with a Type Ib SN (Filippenko et al. 2004; Harutyunyan et al. 2008; Modjaz et al. 2014). No signs of CSM interaction via optical spectroscopy have ever been reported for SN\,2004dk during its photospheric, nebular, or late phases. However, signs of interaction were reported at other wavelengths. The SN's radio luminosity was observed to have rebrightened at $\sim1660$ days (Stockdale et al. 2009) and persisted to at least $\sim1870$ days (Wellons et al. 2012), which indicated that the progenitor mass-loss rate at $\sim100$ yr before explosion was $\dot{M}\approx6.3\times10^{-6}\,{\rm M}_{\odot}$, consistent with red supergiant winds (van Loon et al. 2005). X-ray emission was reported at an epoch of $\sim10$ days (Pooley 2007; Stockdale et al. 2009), yet the properties were typical of SNe~Ib observed at a similar phase (e.g., Soderberg et al. 2008). Optical spectroscopy from post-nebular interaction-dominated phases has never, to our knowledge, been published. However, Vinko et al. (2017) recently reported the detection of late-time narrowband H$\alpha$ excess from SN\,2004dk and other H-poor SNe, motivating our spectroscopic follow-up observations.

We present deep late-time optical spectroscopy of SN\,2004dk, showing that CSM interaction is ongoing and now involves \textit{H-rich} CSM produced centuries before explosion. We also present new optical spectra of SNe\,2014C, 2009ip, and PTF11iqb, showing that strong CSM interaction is persistent in these SNe. We briefly discuss the implication of these observations for stellar evolution and the pathway to stripped-envelope SNe. 

\section{Observations}

SN\,2004dk was observed on 2017 May 29 (UT dates are used throughout this paper) with the Keck-II 10\,m telescope and Deep Imaging Multi-Object Spectrograph (DEIMOS; Faber et al.\ 2003). The date of observation corresponds to an epoch of 4684 days past the discovery date of 2004 Aug. 1 (Graham \& Li 2004). We utilized the 1200~l\,mm$^{-1}$ grating on DEIMOS and the 0\farcs8 slit, which provided a spectral resolving power of $R \equiv \delta \lambda / \lambda \approx4000$. Two exposures of 1500\,s and 1800\,s were obtained at airmass of 1.1 after blindly offsetting $87\farcs4$ W and $21\farcs9$ N from GAIA source ID 4358747869584043520, which has coordinates of $\alpha=245.477757500$\,deg $\delta=-2.278004722$\,deg ($G\approx14.20$ mag). 

Spectra of SN\,2014C were obtained at epochs of 23--1327 days after explosion. Observations were taken using Keck/DEIMOS on 2014 Oct. 2, 2016 Oct. 25, 2017 Aug. 18; and also with the Low-Resolution Imaging Spectrometer (LRIS; Oke et al. 1995) on 2014 July 29 and Nov. 20; and 2015 June 16 and Sep. 16. The LRIS data have a spectroscopic resolving power of $R\approx 2200$. Observations of SN\,2014C were also made with the Kast spectrograph (Miller \& Stone 1993) on the 3\,m Shane reflector at Lick Observatory on the following dates: 2014 Jan. 22; June 20, 29; Aug. 25; and Sep. 3, 28. Kast has moderate resolution ($R \approx 900$ and 1400 for the red and blue sides, respectively) and covered a large wavelength range of 3436--9920\,{\AA}. 

Spectra of SN\,2009ip and PTF\,11iqb were obtained with Keck/DEIMOS on 2017 Aug. 18, utilizing the same instrument settings as for SN\,2004dk, described above. The observation date for SN\,2009ip corresponds to 1883 days past the onset of the 2012B phase, interpreted as the explosion date by Smith, Mauerhan, \& Prieto (2014). For PTF\,11iqb, the observation date corresponds to 2218 days past first detection (Smith et al. 2015). Three exposures were obtained for both objects, totaling 3600\,s at airmass values of 1.55 for SN\,2009ip and 1.15 for PTF\,11iqb.

All spectra were flat fielded and wavelength calibrated using continuum and line lamp exposures. Flux calibration was performed using spectra of standard stars observed on the same night and at similar values of airmass. IRAF\footnote{IRAF is distributed by the National Optical Astronomy Observatory, which is operated by the Association of Universities for Research in Astronomy (AURA) under a cooperative agreement with the US National Science Foundation.} routines were used to process the data. The spectra of SN\,2004dk were Doppler corrected using the redshift of $z=0.005236$ for the host galaxy NGC\,6118 . Luminosity values were estimated using the cosmic distance of  21.05 Mpc (mean value from NASA Extragalactic Database). A distance of 14.7 Mpc was adopted for SN\,2014C (Milisavljevic et al. 2015).

\begin{figure}
\includegraphics[width=3.3in]{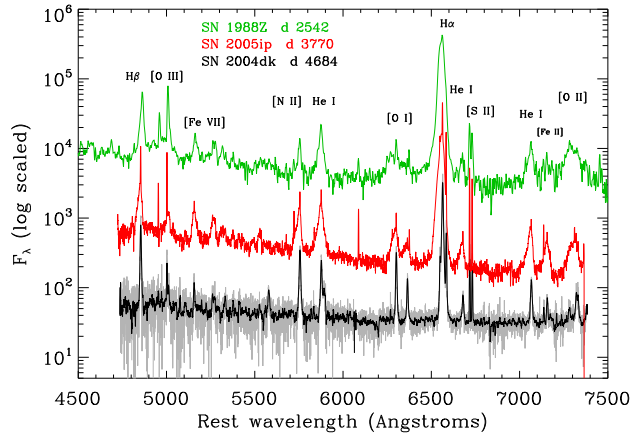}
\caption{Optical spectrum of SN\,2004dk (grey; black smoothed to 20\,{\AA}), compared to spectra of SN\,1998Z and SN\,2005ip (data reproduced from Smith et al. 2017).}
\label{fig:spec_04dk_compare}
\end{figure}

 \section{Results}
 \subsection{SN\,2004dk}
The late-time spectrum of SN\,2004dk is shown in Figure\,\ref{fig:spec_04dk_compare}, along with the previously reported late-time spectra of SN\,2005ip and SN\,1988Z (Smith et al. 2017) for comparison. The spectrum exhibits prominent intermediate-width emission lines of H$\alpha$, He\,{\sc i}, [O\,{\sc i}],  [O\,{\sc ii}], and [Fe\,{\sc ii}], and we attribute these lines to the SN, specifically, CSM gas energized by the forward shock (e.g., see Chugai \& Danziger 1994). At the blue end of the spectrum, H$\beta$, [O\,{\sc iii}], and possible emission lines from [Fe\,{\sc vii}] are also detected. A narrower component of H$\alpha$ and narrow lines of [N\,{\sc ii}] are superimposed on the intermediate-width H$\alpha$ lines; these components are probably associated with host-galaxy H\,{\sc ii} regions (strong spatially-extended background emission was obvious in the two-dimensional spectral images). The narrow features have full width at half-maximums intensity (FWHM) values of $\sim120$--130~km~s$^{-1}$, approximately consistent with the widths of comparison-lamp and night-sky lines, so they are likely to be intrinsically narrower than we can measure. Narrow lines from the [S\,{\sc ii}] doublet at $\lambda\lambda$6717, 6731 are probably associated with a background H\,{\sc ii} region as well. However, it is important to note that in the comparable case of SN\,2005ip, narrow features of [N\,{\sc ii}] and [S\,{\sc ii}] are possibly attributable to outer unshocked low-density CSM photoionized by the SN (Smith et al. 2017), but since the spatially-extended background emission is so strong for SN\,2004dk we strongly suspect that the narrow lines are dominated by H\,{\sc ii} regions rather than CSM.

\begin{figure}
\includegraphics[width=3.3in]{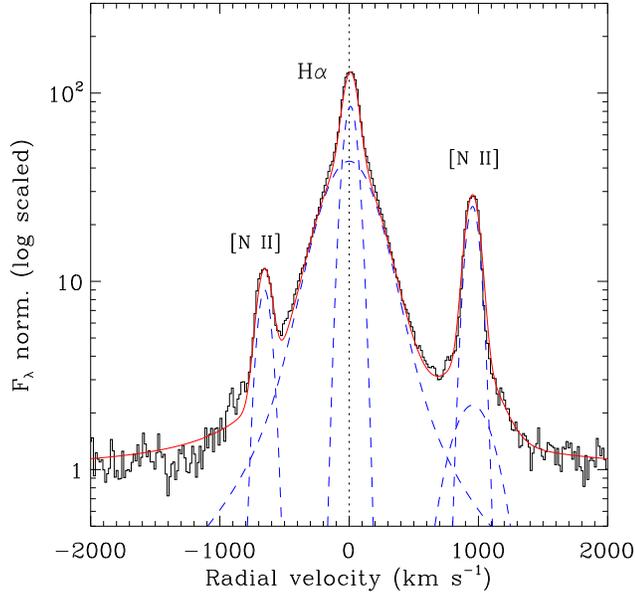}
\caption{H$\alpha$ profile of SN\,2004dk from day 4684 (black). The profile is matched by a superposition (red curve) of several individual components (blue dashed curves), including an intermediate-width Lorentzian profile (400 km s$^{-1}$) of H$\alpha$ and narrow Gaussian profiles (FWHM = 130 km s$^{-1}$) of H$\alpha$ and [N\,{\sc ii}]. }
\label{fig:spec_04dk_ha}
\end{figure}

A more detailed view of the H$\alpha$ profile in SN\,2004dk is shown in Figure\,\ref{fig:spec_04dk_ha}. The spectral resolution is just high enough to distinguish the contributions from intermediate-width features from the SN and narrow features of H$\alpha$ and [N\,{\sc ii}] from the host H\,{\sc ii} region (or photoionized CSM). We modeled the intermediate-width profile using a Voigt function composed of a Gaussian component and a broader Lorentzian component. The main H$\alpha$ feature has $\rm FWHM\approx400$\,km\,s$^{-1}$. The FWHM of the narrow H$\alpha$ and [N\,{\sc ii}] components is $\sim$130\,km\,s$^{-1}$. We note that in order to obtain a satisfactory fit to the entire profile, it was necessary to also include an intermediate-width component of the [N\,{\sc ii}] profile on the positive-velocity side of H$\alpha$, having the same width parameter as the main H$\alpha$ feature. We also tried a single Lorentzian profile fit to the [N\,{\sc ii}] feature, but the results were not satisfactory. 

The integrated flux of the intermediate-width component of our fit to the H$\alpha$ profile is $\sim3.6\times10^{-14}$\,erg\,s$^{-1}$\,cm$^{-2}$, which implies a line luminosity of $L_{\rm H\alpha}\approx2.5\times10^{39}$\,erg\,s$^{-1}$, given the mean metric distance of 20.76 Mpc and the $R$-band extinction of $A_R=0.342$\,mag (both values adopted from NED). The total flux of the entire profile, including the narrow H$\alpha$ and [N\,{\sc ii}] lines from the H\,{\sc ii} region, is $\sim6.3\times10^{-14}$\,erg\,s$^{-1}$\,cm$^{-2}$, implying a total profile luminosity of $\sim4.5\times10^{39}$\,erg; this is nearly 0.5 dex higher than the earlier narrowband imaging-based measurement from Vinko et al. (2017), which would have included the combined flux from all of the profile components. Therefore, it appears possible that the H$\alpha$ flux has increased in just the last $\sim$1000 days.

 \subsection{SNe\,2014C, 2009ip, and PTF11iqb}

\begin{figure}
\includegraphics[width=3.45in]{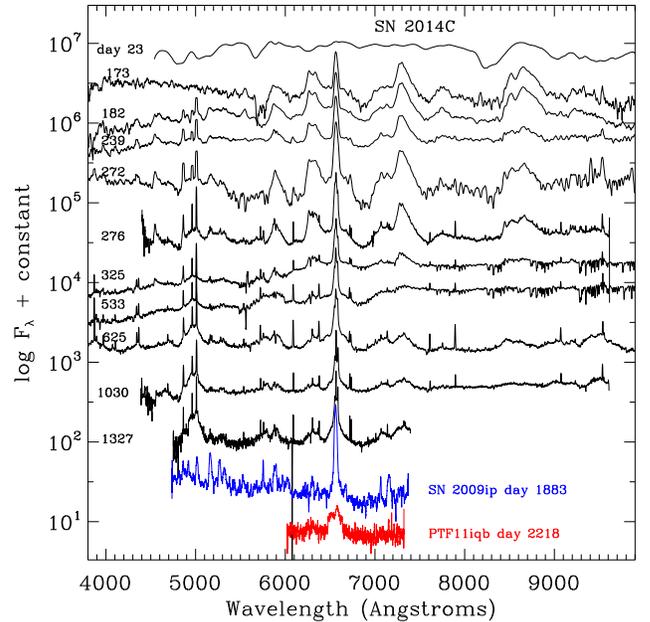}
\caption{Spectral series of SN~2014C (black) between day 23 and 1327, and single late-time spectra of SN\,2009ip on day 1883 (blue) and PTF11iqb on day 2218 (red).}
\label{fig:sn2014c}
\end{figure}

The spectral evolution of SN\,2014C is illustrated in Figure\,\ref{fig:sn2014c}. Only the earliest spectrum on day 23 is from the pre-interaction phase, when the SN was classifiable as a hydrogen-deficient Type Ib. All of the later spectra from days 172--1327 were obtained after the SN had transitioned to a H-rich Type IIn (Milisavljevic et al. 2015). We refer the reader to Milisavljevic et al. (2015; their Figure~6) for detailed line identifications, but note that only our spectra include coverage of the Ca\,{\sc ii} IR triplet at redder wavelengths as well. Overall, the spectra exhibit a blend of intermediate-width lines from shocked CSM, and narrow features that could either be the result of unshocked photoionized CSM (Milisavljevic et al. 2015) or from an H\,{\sc ii} region in NGC~7331 near the SN.

We measured the integrated flux of H$\alpha$ in SN\,2014C over the wavelength region 6520--6620\,{\AA}. Using the latest Keck/DEIMOS spectrum from day 1327, we were able to measure the narrow and intermediate-width line components separately. The integrated continuum-subtracted flux of the intermediate-width component on day 1327 is $\sim1.3\times10^{-14}$\,erg\,s$^{-1}$\,cm$^{-2}$. The total flux of the combined narrow H$\alpha$ and [N\,{\sc ii}] components is $L_{\rm narrow}(\rm H\alpha$+[N\,{\sc ii}])\,$=4.9\times10^{-15}$\,erg\,s$^{-1}$. Given the adopted total extinction of $E(B-V)=0.83$ mag and the distance of 14.7 Mpc (Milisavljevic et al. 2015; Margutti et al. 2017), the intermediate-width H$\alpha$ line luminosity on day 1327 is $L_{\rm H\alpha}\approx2.5\times10^{39}$\,erg\,s$^{-1}$.

The integrated H$\alpha$ flux of PTF11iqb on day 2218 is $\sim4.2\times10^{-15}$\,erg\,s$^{-1}$\,cm$^{-2}$, which implies a line luminosity of $L_{\rm H\alpha}\approx1.3\times10^{39}$\,erg\,s$^{-1}$, given the adopted total extinction and distance estimates from Smith et al. (2015). The H$\alpha$ profile exhibits an asymmetric double-peaked morphology, where the redder of the two peaks exhibits the strongest flux. No major change is apparent in the H$\alpha$ line profile since the day 1104 spectrum reported by Smith et al (2015); we measure only a marginal decrease in luminosity since that measurement.

The integrated H$\alpha$ flux of SN\,2009ip  at 1883 days after the onset of the ``2012B" event (note that the 2012B interaction event began 47 days after the explosion date estimated by Smith, Mauerhan, \& Prieto 2014) is $\sim4.4\times10^{-15}$\,erg\,s$^{-1}$\,cm$^{-2}$, which implies a line luminosity of $L_{\rm H\alpha}\approx2.2\times10^{38}$\,erg\,s$^{-1}$, using the same distance as adopted by Mauerhan et al. (2013). The H$\alpha$ profile is asymmetric, exhibiting the same red shoulder reported by Graham et al. (2017) around day 1026. Based on the latest H$\alpha$ flux measurement, there is some indication that the decline rate is slowing.

\begin{figure}
\includegraphics[width=3.4in]{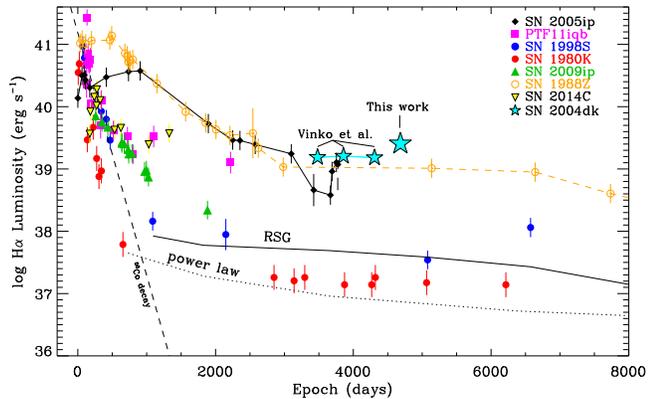}
\caption{Luminosity of H$\alpha$ emission for SN\,2004dk (cyan 5-pointed star), compared to earlier narrow-band measurements from Vinko et al. (2017), and new and previously reported data on SNe\,2014C, PTF11iqb, and 2009ip (Smith et al. 2015; Graham et al. 2017). SNe\,2005ip, 1988S, and 1988Z are also shown (data reproduced from Smith et al. 2017 and Mauerhan \& Smith 2012). Predictions for red-supergiant and power-law CSM models (Chevalier \& Fransson 1994) are shown by the solid black and dotted curves, respectively. The radioactive decay rate of $^{56}$Co is shown as the black dashed line.}
\label{fig:ha_lum_compare}
\end{figure}

Figure\,\ref{fig:ha_lum_compare} illustrates the evolution of H$\alpha$ line luminosity for SN\,2004dk, compared to the long-term trends of other interacting SNe whose H$\alpha$ evolution has been tracked out to late times (Mauerhan \& Smith 2012; Smith et al. 2015, 2017; Graham et al. 2017), including our new measurements of SNe\,2014C, PTF11iqb, and 2009ip. SN\,2004dk exhibits higher H$\alpha$ luminosity than any other interacting SN at a similarly late phase. The possible luminosity increase for the latest epoch is reminiscent of the late-time increase reported for SN\,2005ip by Smith et al. (2017).

The H$\alpha$ evolution of SN\,2014C, starting with our earliest day 173 observation of the H-rich interaction phase, rises to peaks near 276 days and then declines to a plateau beyond day 500, after which it exhibits a luminosity evolution that is similar to that of PTF11iqb. The flattening out of the light curve resembles that of the persistently interacting SNe\,2005ip and 1988Z at similar phases in their evolution. The H$\alpha$ decline rate of SN\,2009ip is initially comparable to those of SNe\,2014C and PTF11iqb up to  $\sim1000$ days; however, by 2000 days SN\,2009ip becomes fainter overall by $\sim1$ dex, and almost as faint as SNe\,1998S and 1980K (Mauerhan \& Smith 2012).

\section{Discussion}

\begin{figure}
\includegraphics[width=3.45in]{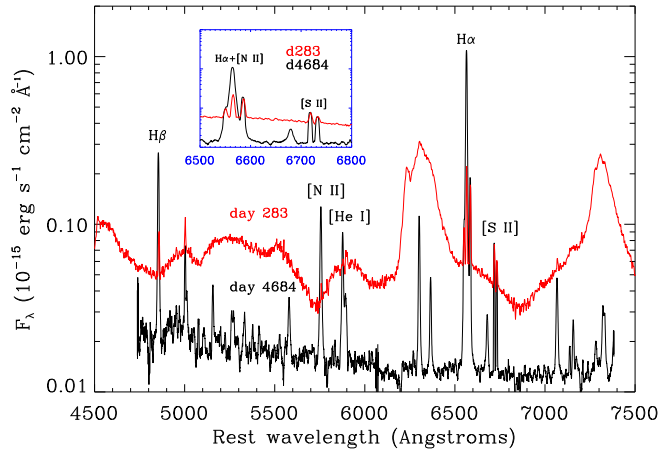}
\caption{Optical spectra of SN\,2004dk on day 4684 (black) compared to day 283 (red) and smoothed to the lower 2~{\AA} resolution of the later. The blue inset panel shows a zoom of the wavelength region 6600--6800~{\AA}.} \label{fig:spec_04dk_epochs}
\end{figure}

Signatures of CSM interaction were not detected in any of the previously reported optical spectra of SN\,2004dk, so it is unclear when interaction with H-rich CSM began. To determine whether the H$\alpha$ interaction signatures seen on day 4684 could have also been present at early times, but washed out from the bright SN emission, we compare with LRIS spectra from day 283 in Figure\,\ref{fig:spec_04dk_epochs} (the day 283 spectrum was used in the analysis of Shivvers et al. 2017, but not displayed in that paper).  The only feature of H$\alpha$ seen on day 283 is a very narrow feature that probably originates in a background host H\,{\sc ii} region; no Balmer emission is obviously attributable to the SN. The flux of the [S\,{\sc ii}] $\lambda\lambda$6717, 6731 doublet from the background [H\,{\sc ii}] region is the same at both epochs (within $\sim5$\%), as is the narrow [N\,{\sc ii}] doublet bracketing H$\alpha$ (the consistency in measured flux for the H\,{\sc ii} region lines provides reassurance that our flux calibration is accurate, and that our measurement of the H$\alpha$ flux on day 4684 is reliable). 

However, H$\alpha$ and H$\beta$ are much stronger on day 4684 relative to day 283, and so is [N\,{\sc ii}] at $\lambda$5755. The comparison indicates that the interaction-powered H$\alpha$ luminosity we observe on day 4684 would have been discernible in the day 283 spectrum, had it been exhibiting the same level of strength. This confirms that the interaction with H-rich CSM was delayed until after day 283. Therefore, from our data alone, one could very roughly approximate that the interaction with H-rich CSM probably began sometime between day 283 and 4684. The narrowband imaging detection of H$\alpha$ line excess from Vinko et al. (2017) further restricts the H-rich interaction onset time to before day $\sim3500$. We therefore suspect that the interaction with H-rich CSM commenced at the same times as the radio rebrightening on day $\sim1660$ reported by Stockdale et al. (2009). Indeed, in the relatable case of SN\,2014C, delayed interaction with H-rich material was also accompanied by an increase in radio luminosity (Margutti et al. 2017). Therefore, using the radio rebrightening at 1660 days as the onset of interaction with H-rich material, and approximating a SN shock velocity of 0.1c, the implied distance to the inner edge of the H-rich CSM in SN\,2004dk is $4\times10^{17}$~cm, or $\sim0.1$ pc. By comparison, the estimated radius to the dense H-rich CSM in SN\,2014C is an order of magnitude smaller, at (3--6) $\times10^{16}$~cm (Milisavjevic et al. 2015; Margutti et al. 2017).

With no optical spectra of SN\,2004dk available from near the earlier interaction-onset time, we are unable to obtain the CSM velocity information necessary to derive an accurate look-back time for the event that created the separated structure of H-rich CSM. However, if we were to hypothetically assume that the CSM had the same outflow velocity as indicated by the line widths of $\sim400$~km~s$^{-1}$ that we measured at late times, then the associated look-back time of the mass-loss event is $\sim320$ yr before explosion. If the cavity was instead inflated by a WR-like wind with a velocity of 1000~km~s$^{-1}$, then the wind would have initiated $\sim125$ yr before explosion. On the longer term, the invigorated interaction at 4684 days implies a distance limit of $<10^{18}$~cm for the CSM density enhancement, with the actual value being dependent on the deceleration of the shock over that timescale, which is probably non-negligible. 

As demonstrated in Figure\,\ref{fig:spec_04dk_compare}, most of same atomic transitions observed in SN\,2004dk are also present in persistently interacting SNe~IIn 2005ip and 1988Z at similarly late times (Smith et al. 2017), although the line widths are narrower in SN\,2004dk (FWHM $\approx400$\,km\,s$^{-1}$). The similarities suggest that the kinematics and chemistry of the CSM in each of these SNe are comparable, and therefore the CSM could have plausibly arisen from similar pre-SN stellar evolutionary processes for massive stars. Like SN\,2005ip and SN\,1988Z, the emission-line widths in SNe\,2004dk, 2014C, and 2009ip are all indicative of excitation of the CSM gas by  passage of the forward shock, and the endurance of this emission indicates that the shocks from these SNe are still propagating through an extended distribution of CSM. The possible H$\alpha$ luminosity increase in SN\,2004dk is also reminiscent of the late-time fluctuations reported for SN\,2005ip, suggesting that the stellar progenitors of SNe~IIn and stripped-envelope SNe~Ib can experience significant fluctuations in their pre-SN mass-loss rates.

The endurance of the forward shock in SN\,2004dk, as in the cases of SNe\,2005ip and 1988Z (Smith et al. 2017), seems to favor a strong, persistent wind before the associated mass-loss transition that created the cavity occurred; and the same reasoning seems appropriate in the case of SN\,2014C. Interestingly, it has been suggested that red supergiants can experience pulsations driven by the partial ionization of the H envelope, and if the pulsation growth rates becomes high enough, a ``superwind" can be triggered (Heger et al. 1997). Simulations of this process have shown that pulsationally driven superwinds could be marked by a runaway increase in mass-loss rate, expelling a large fraction of the H envelope, followed by a sudden decrease in mass-loss rate, and all on a $\lesssim10^3$ year timescale before the SN (Yoon \& Cantiello 2010). We speculate that this process might explain the delayed CSM interaction in SN\,2004dk and SN\,2014C, creating rarefied cavities rimmed by dense CSM. If, after losing a substantial fraction of its H envelope, the progenitor of SN\,2004dk transitioned from a red supergiant (with a slow H-rich wind with mass-loss rate of $\dot{M}\approx6.3\times10^{-6}\,M_{\odot}$, as indicated by radio data on day 1660 from Stockdale et al. (2009)), into a H-deficient WR star with a rarefied hypersonic wind (consistent with the early Type Ib spectrum), then a bubble with a dense rim would have developed as the slower H-rich material was swept up (e.g., Freyer et al. 2006). Since the interaction luminosity is sensitive to the difference in velocity between the SN and CSM outflow, interaction signatures can be relatively weak and difficult to detect while the shock propagates through the fast rarefied CSM of the WR phase, becoming more pronounced after the inner edge of the slower dense wind of the red supergiant phase is reached. 

Alternatively, the progenitors of these SNe could have also experienced discrete eruptions (Smith \& Owocki 2006; Smith 2014); if the momentum of the ejected  mass is sufficiently high, such an eruption could carve out a cavity within the more extended wind material. Indeed, spectroscopic transitions between blue-supergiant and WR phases have been observed in evolved massive-star systems, such as HD\,5980, and punctuated by intermittent LBV eruptions (Koenigsberger et al. 2010). The radial density profile of the CSM surrounding such systems could be quite complex, giving rise to highly variable interaction signatures after the SN occurs. 

Stripped-envelope SNe of Type Ib, which both SNe\,2004dk and 2014C were at the time of explosion, have long been suspected to stem from interacting binary systems. Mass loss initiated by Roche-lobe overflow (RLOF) and common-envelope evolution (CEE) could therefore be a natural pathway for creating displaced H-rich CSM in these SNe. RLOF and CEE could be linked to advanced shell-burning episodes during pre-SN evolution, as stars could swell in radius or pulsate in response to shell burning and in turn initiate stellar interactions, mass-transfer episodes, and even discrete eruptions (Shiode \& Quataert 2014; Smith \& Arnett 2014; Smith 2014, and references therein). 

Whatever the origin of the pre-SN mass loss in SN\,2004dk, it is likely to be a detectable X-ray source, given the unprecedentedly high H$\alpha$ luminosity for its age, and we are planning such X-ray observations.  As in the relatable cases of SN\,2005ip (Smith et al. 2017) and SN\,2014C (Margutti et al. 2017), X-ray data will yield additional insights into the physical parameters of the shock-energized CSM and further constrain its origin.

\section*{Acknowledgments}
Some of the data presented herein were obtained at the W. M. Keck
Observatory, which is operated as a scientific partnership among the
California Institute of Technology, the University of California, and
NASA; the observatory was made possible by the generous financial
support of the W. M. Keck Foundation. We thank the staffs of Lick and
Keck Observatories for their assistance with the observations.  We
thank Brad Cenko, Thomas de Jaeger, Ori Fox, Pat Kelly, Brad Tucker, 
and Heechan Yuk for help with some of the observations and reductions.
Research at Lick Observatory is partially supported by a generous gift
from Google.  The work of A.V.F.'s supernova group at UC Berkeley has
been generously supported by the TABASGO Foundation, the Christopher
R. Redlich Fund, and the Miller Institute for Basic Research in
Science (U.C. Berkeley).


\begin{thebibliography}{99}
\bibitem[Begelman \& Sarazin(1986)]{1986ApJ...302L..59B} Begelman, M.~C., \& Sarazin, C.~L.\ 1986, ApJL, 302, L59 
\bibitem[Chevalier \& Fransson(1994)]{1994ApJ...420..268C} Chevalier, R.~A., \& Fransson, C.\ 1994, ApJ, 420, 268 
\bibitem[Chugai \& Danziger(1994)]{1994MNRAS.268..173C} Chugai, N.~N., \& Danziger, I.~J.\ 1994, MNRAS, 268, 173
\bibitem[Dwarkadas(2007)]{2007ApJ...667..226D} Dwarkadas, V.~V.\ 2007, ApJ, 667, 226 
\bibitem[Dwarkadas(2005)]{2005ApJ...630..892D} Dwarkadas, V.~V.\ 2005, ApJ, 630, 892 
\bibitem[Faber et al.(2003)]{2003SPIE.4841.1657F} Faber, S.~M., Phillips, A.~C., Kibrick, R.~I., et al.\ 2003, SPIE, 4841, 1657 
\bibitem[Filippenko et al.(2004)]{2004IAUC.8404....1F} Filippenko, A.~V., Ganeshalingam, M., Serduke, F.~J.~D., \& Hoffman, J.~L.\ 2004, IAUC, 8404, 1 
\bibitem[Freyer et al.(2006)]{2006ApJ...638..262F} Freyer, T., Hensler, G., \& Yorke, H.~W.\ 2006, ApJ, 638, 262 
\bibitem[Graham et al.(2017)]{2017MNRAS.469.1559G} Graham, M.~L., Bigley, A., Mauerhan, J.~C., et al.\ 2017, MNRAS, 469, 1559 
\bibitem[Graham \& Li(2004)]{2004IAUC.8377....2G} Graham, J., \& Li, W.\ 2004, IAUC, 8377, 2 
\bibitem[Harutyunyan et al.(2008)]{2008A&A...488..383H} Harutyunyan, A.~H., Pfahler, P., Pastorello, A., et al.\ 2008, A\&A, 488, 383 
\bibitem[Heger et al.(1997)]{1997A&A...327..224H} Heger, A., Jeannin, L., Langer, N., \& Baraffe, I.\ 1997, A\&A, 327, 224 
\bibitem[Heikkil{\"a} et al.(2016)]{2016MNRAS.457.1107H} Heikkil{\"a}, T., Tsygankov, S., Mattila, S., et al.\ 2016, MNRAS, 457, 1107 
\bibitem[Koenigsberger et al.(2010)]{2010AJ....139.2600K} Koenigsberger, G., Georgiev, L., Hillier, D.~J., et al.\ 2010, AJ, 139, 2600 
\bibitem[Leonard et al.(2000)]{2000ApJ...536..239L} Leonard, D.~C., Filippenko, A.~V., Barth, A.~J., \& Matheson, T.\ 2000, ApJ, 536, 239 
\bibitem[Maeda et al.(2008)]{2008Sci...319.1220M} Maeda, K., Kawabata, K., Mazzali, P.~A., et al.\ 2008, Science, 319, 1220 
\bibitem[Mauerhan \& Smith(2012)]{2012MNRAS.424.2659M} Mauerhan, J., \& Smith, N.\ 2012, MNRAS, 424, 2659 
\bibitem[Mauerhan et al.(2013)]{2013MNRAS.430.1801M} Mauerhan, J.~C., Smith, N., Filippenko, A.~V., et al.\ 2013, MNRAS, 430, 1801 
\bibitem[Mauerhan et al.(2014)]{2014MNRAS.442.1166M} Mauerhan, J., Williams, G.~G., Smith, N., et al.\ 2014, MNRAS, 442, 1166  
\bibitem[Margutti et al.(2017)]{2017ApJ...835..140M} Margutti, R., Kamble, A., Milisavljevic, D., et al.\ 2017, ApJ, 835, 140 
\bibitem[Milisavljevic et al.(2015)]{2015ApJ...815..120M} Milisavljevic, D., Margutti, R., Kamble, A., et al.\ 2015, ApJ, 815, 120 
\bibitem[Modjaz et al.(2014)]{2014AJ....147...99M} Modjaz, M., Blondin, S., Kirshner, R.~P., et al.\ 2014, AJ, 147, 99 
\bibitem[Niedzielski \& Skorzynski(2002)]{2002AcA....52...81N} Niedzielski, A., \& Skorzynski, W.\ 2002, AcA, 52, 81 
\bibitem[Ofek et al.(2014)]{2014ApJ...789..104O} Ofek, E.~O., Sullivan, M., Shaviv, N.~J., et al.\ 2014, ApJ, 789, 104 
\bibitem[Patat et al.(2004)]{2004IAUC.8379....3P} Patat, F., Pignata, G., Benetti, S., \& Aceituno, J.\ 2004, IAUC, 8379, 3 
\bibitem[Pastorello et al.(2007)]{2007Natur.447..829P} Pastorello, A., Smartt, S.~J., Mattila, S., et al.\ 2007, Nature, 447, 829 
\bibitem[Podsiadlowski et al.(1992)]{1992ApJ...391..246P} Podsiadlowski, P., Joss, P.~C., \& Hsu, J.~J.~L.\ 1992, ApJ, 391, 246 
\bibitem[]{}Pooley, D. 2007, in AIP Conf. Ser. 937, Supernova 1987A: 20 Years After: Supernovae and Gamma-Ray Bursters, ed. S. Immler, K. Weiler, \& R. McCray (Melville, NY: AIP), 381
\bibitem[Foley et al.(2007)]{2007ApJ...657L.105F} Foley, R.~J., Smith, N., Ganeshalingam, M., et al.\ 2007, ApJL, 657, L105 
\bibitem[Shiode \& Quataert(2014)]{2014ApJ...780...96S} Shiode, J.~H., \& Quataert, E.\ 2014, ApJ, 780, 96 
\bibitem[Shivvers et al.(2017)]{2017MNRAS.471.4381S} Shivvers, I., Zheng, W., Van Dyk, S.~D., et al.\ 2017, MNRAS, 471, 4381 
\bibitem[Smith et al.(2017)]{2017MNRAS.466.3021S} Smith, N., Kilpatrick, C.~D., Mauerhan, J.~C., et al.\ 2017, MNRAS, 466, 3021  
\bibitem[Smith et al.(2015)]{2015MNRAS.449.1876S} Smith, N., Mauerhan, J.~C., Cenko, S.~B., et al.\ 2015, MNRAS, 449, 1876 
\bibitem[Smith et al.(2014)]{2014MNRAS.438.1191S} Smith, N., Mauerhan, J.~C., \& Prieto, J.~L.\ 2014, MNRAS, 438, 1191 
\bibitem[Smith(2014)]{2014ARA&A..52..487S} Smith, N.\ 2014, ARA\&A, 52, 487 
\bibitem[Smith et al.(2009)]{2009ApJ...695.1334S} Smith, N., Silverman, J.~M., Chornock, R., et al.\ 2009, ApJ, 695, 1334 
\bibitem[Smith \& Arnett(2014)]{2014ApJ...785...82S} Smith, N., \& Arnett, W.~D.\ 2014, ApJ, 785, 82 
\bibitem[Smith \& Owocki(2006)]{2006ApJ...645L..45S} Smith, N., \& Owocki, S.~P.\ 2006, ApJL, 645, L45 
\bibitem[Stockdale et al.(2009)]{2009CBET.1714....1S} Stockdale, C.~J., Heim, M.~S., Vandrevala, C.~M., et al.\ 2009, CBET, 1714, 1 
\bibitem[van Loon et al.(2005)]{} van Loon, J. T., Cioni, M. R. L., Zijlstra, A. A., \& Loup, C. 2005, A\&A, 438, 273
\bibitem[Vinko et al.(2017)]{2017ApJ...837...62V} Vinko, J., Pooley, D., Silverman, J.~M., et al.\ 2017, ApJ, 837, 62 
\bibitem[Wellons et al.(2012)]{2012ApJ...752...17W} Wellons, S., Soderberg, A.~M., \& Chevalier, R.~A.\ 2012, ApJ, 752, 17 
\bibitem[Wheeler \& Levreault(1985)]{1985ApJ...294L..17W} Wheeler, J.~C., \& Levreault, R.\ 1985, ApJL, 294, L17 
\bibitem[Yoon \& Cantiello(2010)]{2010ApJ...717L..62Y} Yoon, S.-C., \& Cantiello, M.\ 2010, ApJL, 717, L62 


\end{thebibliography}
\end{document}